\documentclass[12pt]{article}
\usepackage{aas_macros}
\usepackage{times}
\usepackage{geometry}
\usepackage[utf8]{inputenc}
\usepackage{enumitem,amssymb}
\usepackage{ragged2e}
\usepackage{graphicx}
\usepackage{fancyhdr}
\usepackage{hyperref}
\usepackage{tabularx}
\usepackage[
    natbib=true,
    style=numeric,
    sorting=none
]{biblatex}
\usepackage{color, colortbl}
\newlist{thematic}{itemize}{8}
\setlist[thematic]{label=$\square$}
\usepackage{pifont}

\newcommand\farcs{\mbox{$.\!\!^{\prime\prime}$}}%

\begin{document}
\raggedright
\huge{The HIP 54515 b Superjovian Planet as
an Early, Critical Look at the Roman Coronagraph's Performance in the Faint Target Star, Small IWA Limit}
\linebreak
\large

Thayne Currie$^{1,2}$,Yiting Li$^{3}$, Brianna Lacy$^{4}$, Mona El Morsy$^{1}$, Masayuki Kuzuhara$^{5}$, Naoshi Murakami$^{5}$, Danielle Bovie$^{1}$

1 Department of Physics and Astronomy, University of Texas-San Antonio, San Antonio, TX, USA \\
2 Subaru Telescope, National Astronomical Observatory of Japan, 
Hilo, HI, USA\\
3 Department of Astronomy, University of Michigan, Ann Arbor, MI, USA\\
4 Department of Astronomy \& Astrophysics, University of California-Santa Cruz, Santa Cruz, CA, USA\\
5 National Astronomical Observatory of Japan, 2-21-1 Osawa, Mitaka, Tokyo 181-8588, Japan

\justify{
\textbf{Abstract:}  
The Roman Coronagraph's capabilities in the faint star, small IWA limit has enormous scientific (programmatic) impacts.  Testing its performance in this limit provides a first look at challenges that may be encountered with the \textit{Habitable Worlds Observatory} in imaging rocky planets around the nearest K and M stars.  We propose such a rigorous test with the HLC/575nm targeting a newly-discovered superjovian planet HIP 54515 b, whose predicted contrast is $\sim$4.7 $\times$10$^{-8}$--2.5 $\times$10$^{-7}$.  The companion lies close to the coronagraph IWA (well interior to the TTR5 performance region) and orbits a V = 6.8 star, near the limit for which the coronagraph may yield deep contrasts.  Multiple reference stars are available that will further test CGI's performance as a function of $\Delta$ pitch angle to assess how the telescope's thermal environment impacts contrasts.}

\pagebreak
\noindent \textbf{Type of observation:} \\$\boxtimes$   Technology Demonstration\\
$\square$ Scientific Exploration\\\\

\noindent \textbf{Scientific / Technical Keywords:}  
high contrast performance

\noindent \textbf{Required Detection Limit:}  
\begin{tabular}{| c | c | c | c | c |}
\hline
$\geq$10$^{-5}$ & 10$^{-6}$ & 10$^{-7}$ & 10$^{-8}$ & 10$^{-9}$ \\ \hline
 & & x & & \\ \hline
\end{tabular}

\vspace{0.5cm}
\textbf{Roman Coronagraph Observing Mode(s):} 

\begin{tabular}{| c | c | c | c | c |}
\hline
Band &   Mode & Mask Type & Coverage & Support \\ \hline \hline

1, 575 nm &   Narrow FoV & Hybrid Lyot & 360$^{\circ}$ & Required (Imaging) \\
 & Imaging &  &  &  \\ \hline 
\end{tabular}

\justify{
\begin{center}
\begin{tabular}{| c | c | c | c | c |}
\hline
Name &  host star & detection & separation (") & description of target \\
  & V mag. & limit & (or extent)  & \\ \hline \hline
  HIP 54515 b & 6.8 & 4.7$\times$10$^{-8}$ & 0.23 & self-luminous exoplanet\\
  \hline
\end{tabular}
\end{center}

}
\pagebreak
\begin{figure}[!ht]
    \includegraphics[width=0.4\textwidth,clip]{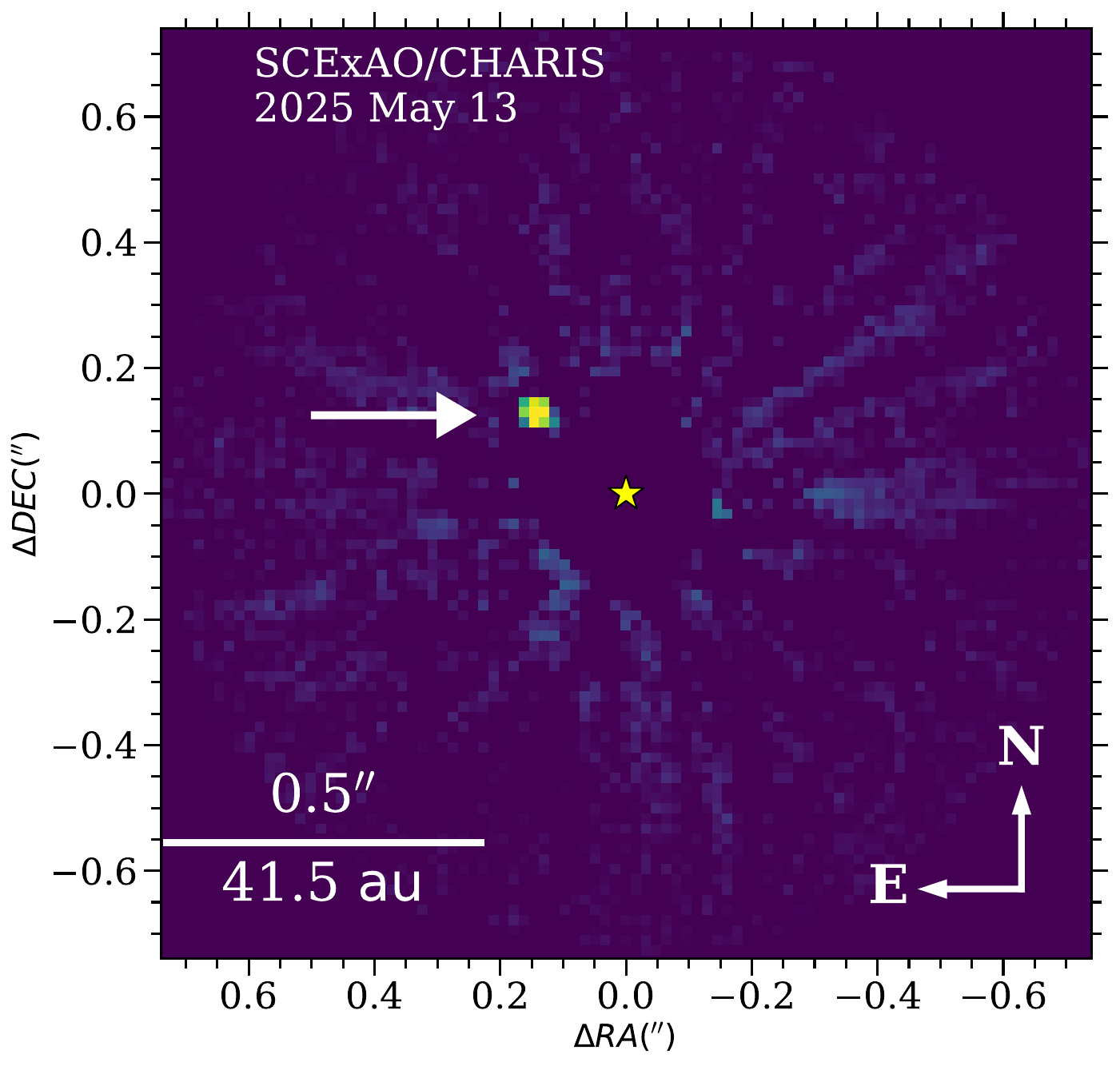}
     \includegraphics[width=0.6\textwidth,clip]{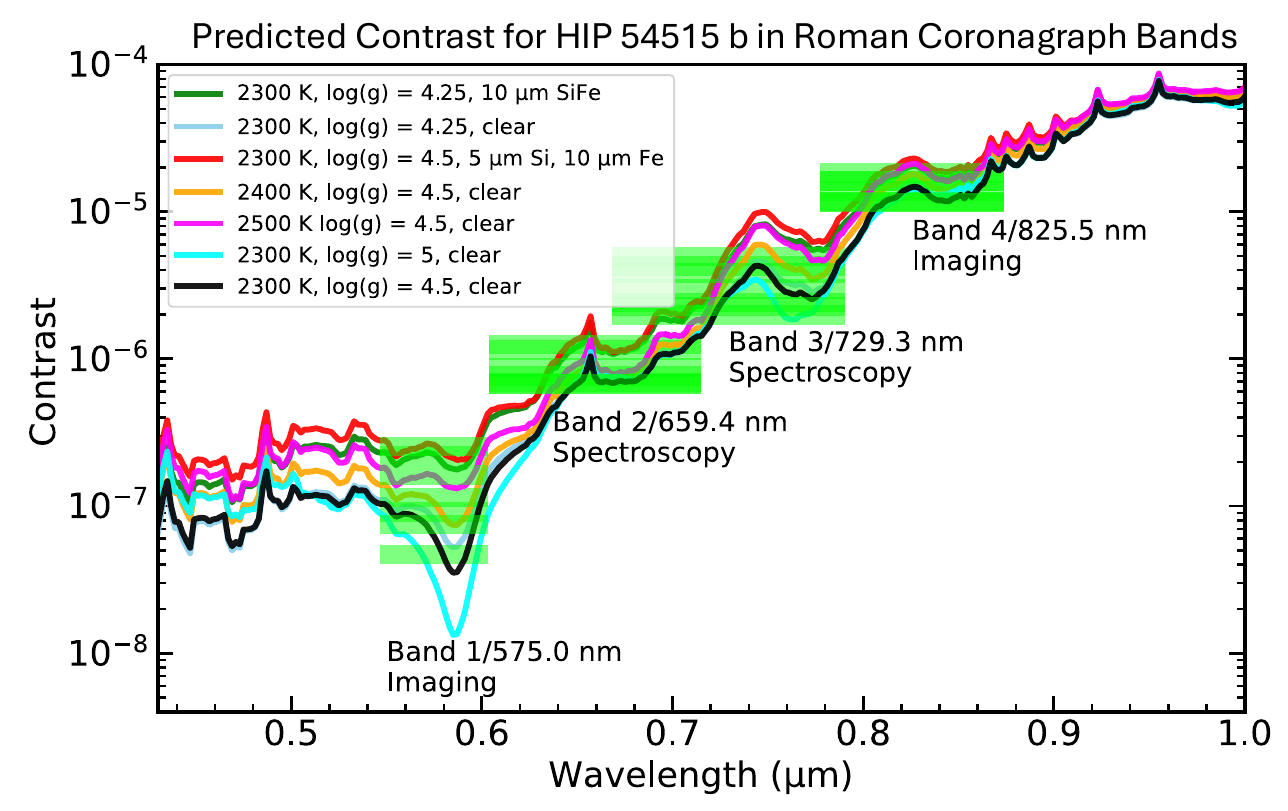}
    \vspace{-0.4in}
    \caption{(left) Latest-epoch detection of HIP 54515 b with SCExAO/CHARIS \citep{Jovanovic2015,Groff2016} at an angular separation of 0\farcs{}19 from the OASIS survey.   The target was selected by the same strategy used to discover and precisely characterize the HIP 99770 b planet and other planets/brown dwarfs by both imaging and astrometry \citep{Currie2023a,Currie2023b,Kuzuhara2022,Franson2023,Tobin2024,ElMorsy2025}.   Orbital modeling suggests that it will lie at $\rho =$ 0\farcs{}2--0\farcs{}24 during the CGI observation phase (e.g., 0\farcs{}23 in January 2028).
    (right) HIP 54515 b's predicted contrast in the four main CGI passbands from updated versions of the \citeauthor{LacyBurrows2020} models.  Modeling its CHARIS spectra and NIRC2 photometry yield a temperature of $\sim$2400 $\pm$ 100 $K$.   At 575 nm, it then has a contrast of $\sim$4.7 $\times$10$^{-8}$ to 2.5 $\times$ 10$^{-7}$.  
    }
\end{figure}
\justify{
\textbf{\Large Anticipated Technology / Science Objectives:}
The Roman Coronagraph (CGI) may be the first high-contrast instrument capable of imaging nearby reflected-light Jupiter-analogue planets ($\Delta$F $\approx$10$^{-9}$) through advanced wavefront control and coronagraphy: a key milestone on the road to imaging an exo-Earth \citep{Kasdin2020,Currie2023b}.  It focuses on achieving this performance on bright stars ($V$ $\lesssim$ 5) and at $\approx$ 6--9 $\lambda$/D.

CGI's performance 1) on fainter stars and 2) just outside its inner working angle (IWA; $\sim$3 $\lambda$/D) is also critically important as a technology demonstration. 
CGI should deliver slightly shallower contrasts for stars as faint as V = 7, but its actual performance degradation from V = 5 to V = 7 is unconstrained.   
Rigorously assessing CGI's performance in the photon-starved, small IWA limit would yield valuable constraints on which scientific programs are feasible in an extended mission and critical input for the \textit{Habitable Worlds Observatory}'s performance in similar situations (e.g. imaging Earths around the nearest K and early M stars).

The best and most striking way to assess CGI's performance in the low-stellar-flux, small-separation regime (e.g. 0\farcs{}15--0\farcs{}25) is to reimage a known, cool substellar companion found from ground-based infrared imaging.   However, until recently, the peer-reviewed literature did not identify any imaged substellar companion orbiting a V = 5--7 star at this separation with a predicted contrast near the TTR5 performance baseline (10$^{-7}$)\footnote{E.g. $\beta$ Pic bc are undetectable due to inner working angle limits and system's debris disk.  Updates to leading published models suggest that HD 206893 B will have a 575 nm contrast $\le$ 10$^{-8}$: it is thus a risky target \citep{LacyBurrows2020}.}.

We propose to investigate CGI's performance in this regime by reimaging HIP 54515 b, a superjovian planet discovered from the Observing Accelerators with SCExAO Imaging Survey (PI T. Currie; \citealt{ElMorsy2024}) funded by NASA to identify targets suitable for the CGI tech demo phase \citep{Currie2025}(Fig. 1).  The latest epoch of SCExAO/CHARIS data resolves HIP 54515 b at 0\farcs{}19: dynamical modeling constrains its 2027--2028 separation to  $\rho$ $\sim$ 0\farcs{}22--0\farcs{}23.  Updated atmospheric models\citep{LacyBurrows2020} suggest that its 575 nm contrast is comparable to that needed to fulfill TTR5.

This program is more technically challenging than that one separately written to achieve TTR5 by reimaging a companion around a V $\sim$ 5 star, as HIP 54515 is a factor of 5 fainter.  However, it provides critical input for science programs that may be considered in an extended observing phase.  Therefore, this observation is best executed during the observation phase soon after TTR5 is achieved on a different target.
}


\textbf{\Large Observing Description:}
The program goal requires Band 1/575 nm imaging of HIP 54515 with the Hybrid Lyot Coronagraph to re-detect its planet.  To directly compare the results to proposed observations of the OASIS discovery target to fulfill TTR5, the observations should follow the ``standard typical observing sequence" described on page 36 of the most recent CGI white paper slide deck that enables both angular and reference star differential imaging (ADI, RDI)\footnote{Coronagraph$\_$CPP$\_$WPoverview2025$\_$8July2025}.  Each "visit" consists of dark hole digging on a bright nearby PSF reference, PSF reference observations, two sets of +/- telescope rolls ($\Delta$$\theta$= +/- 15$^{o}$) on the target star, followed by a second set of PSF reference observations.   

HIP 54515 b's predicted angular separation during the latter parts of the observation phase (e.g. Jan. 2028) is $\sim$ 0\farcs{}22--0\farcs{}23, or $\sim$ 1.5 $\lambda$/D from the CGI inner working angle \citep{Currie2025}.  The primary star is $\sim$ 5.2 times fainter than the nominal TTR5 brightness benchmark (V = 5).  Atmospheric model comparisons to HIP 54515 b's infrared spectrum suggest a 575 nm contrast of $\sim$4.7 $\times$10$^{-8}$ to 2.5 $\times$ 10$^{-7}$ \citep{Currie2025}.   Thus, detecting HIP 54515 at SNR $>$ 11 will demonstrate TTR5-level (10$^{-7}$) contrasts on stars near 7th magnitude in V band at CGI's inner working angle.

HIP 54515 has at least two credible candidate PSF reference stars from the CPP team vetting list: HIP 54872 ($\sim$21$^{o}$ away) and HIP 57632 ($\beta$ Leo, $\sim$19$^{o}$ away) (Fig. 2, left).  Keepout maps (not shown) show that HIP 57632 can only be observed with a pitch angle difference of 5--10$^{o}$ from HIP 54515's pitch angle.  However, HIP 57632 has two $\sim$ 70-day periods (centered on $\sim$100 and 270 days) where it will have $\Delta$ Pitch Angle $<$ 3$^{o}$.  Thus, two sets of observations, one per PSF reference, will probe CGI's low target flux limit over the nominal and wider $\Delta$ pitch angles to assess how performance degrades due to different thermal environments.



\textbf{Estimate of Time Needed}:
For a contrast of 4.7$\times$10$^{-8}$, Corgi-ETC predicts on-source observing times of 2.35 hrs and 5.96 hrs to achieve 10$^{-7}$ 5-$\sigma$ contrasts for the optimistic and conservative performance scenarios, respectively (Fig. 2). 
Adopting the CGI primer efficiency estimate (24 hours clock time for 14 hours observing time),  for each target+PSF reference combination we then estimate a nominal clock time of 4.03 hrs and 10.22 hours for the optimistic and conservative scenarios.  The entire program (HIP 54515 + HIP 57632, HIP 54515 + HIP 54872) can thus be conducted within 14.3 hours.

\vspace{0.2cm}






\begin{figure}[!h]
 \includegraphics[width=0.5\textwidth,clip]{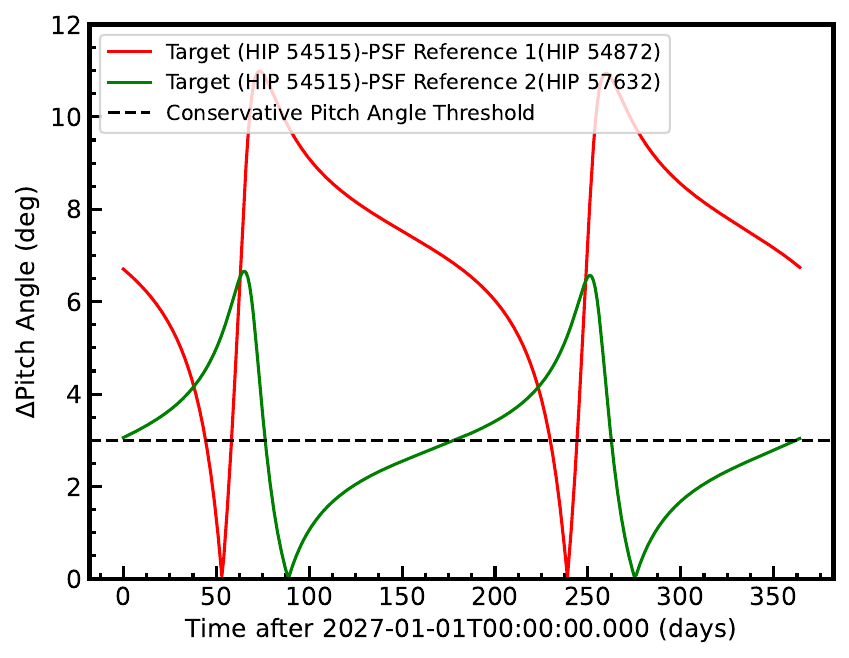}
      \includegraphics[width=0.49\textwidth,clip]{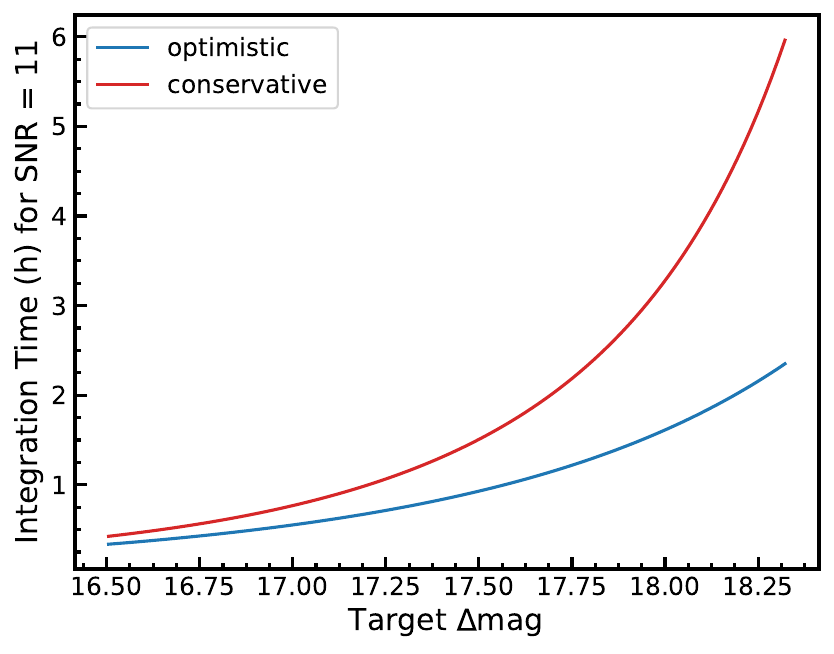}
    \vspace{-0.2in}
    \caption{(left) Relative Pitch Angle between HIP 54515 and two potential reference stars during the calendar year of 2027.  Combined with keepout maps (not shown), the second PSF reference (HIP 57632) would allow observations with a $\Delta$ Pitch Angle less than 3$^{o}$.  (right) Integration time needed to achive a SNR = 13 detection of HIP 54515 b vs. HIP 54515 b contrast at 575 nm for conservative and optimistic CGI performances.  A SNR = 11 detection for our worst-cast contrast scenario (4.7$\times$10$^{-8}$) would demonstrate TTR5-level contrasts on a star 5 times fainter than the TTR5 limit and near CGI's inner working angle.  We assume a solar system exozodi level, though HIP 54515 lacks evidence for circumstellar dust.  }
\end{figure}

\pagebreak

\printbibliography

\end{document}